\documentclass{PoS}

\usepackage{graphicx}
\usepackage{amsmath}
\usepackage{amsfonts}
\usepackage{subfigure}
\usepackage{wrapfig}
\usepackage{epsfig}
\usepackage{afterpage,float}
\usepackage{cite}
\usepackage{ifpdf} 

\newcommand{\be}{\begin{equation}}
\newcommand{\ee}{\end{equation}}
\newcommand{\bea}{\begin{eqnarray}}
\newcommand{\eea}{\end{eqnarray}}

\def\half{{\textstyle{1\over2}}}

\def\bar{\overline}

\def\half{{\scriptstyle \raise.15ex\hbox{${1\over2}$}}}

\newcommand{\beq}{\begin{equation}}
\newcommand{\eeq}{\end{equation}}

\newcommand{\real}{\relax{\rm I\kern-.18em R}}

\title{Conformal finite size scaling of twelve fermion flavors}

\ShortTitle{Conformal finite size scaling of twelve fermion flavors}

\author{Zolt\'an Fodor\\
        Department of Physics, University of Wuppertal\\
        Gau$\beta$strasse 20, D-42119, Germany\\
        J\"ulich Supercomputing Center, Forschungszentrum\\
        J\"ulich, D-52425 J\"ulich, Germany\\
        Email: \email{fodor@bodri.elte.hu}}

\author{Kieran Holland\\
        Albert Einstein Center for Fundamental Physics, Institute for
        Theoretical Physics, \\
        Bern University, Sidlerstrasse 5, CH-3012 Bern, Switzerland\\
      Department of Physics, University of the Pacific,
        3601 Pacific Ave, Stockton CA 95211, USA\\
        Email: \email{kholland@pacific.edu}}

\author{Julius Kuti\\
        Department of Physics 0319, University of California, San Diego\\
        9500 Gilman Drive, La Jolla, CA 92093, USA\\
        E-mail: \email{jkuti@ucsd.edu}}

\author{D\'aniel N\'ogr\'adi\\
        Institute for Theoretical Physics, E\"otv\"os University\\
        H-1117 Budapest, Hungary\\
        Email: \email{nogradi@bodri.elte.hu}}

\author{Chris Schroeder\\
        Physical Sciences Directorate, Lawrence Livermore National Laboratory\\
        Livermore, California 94550, USA\\
        E-mail: \email{schroeder10@llnl.gov}}
        
\author{\speaker{Chik Him Wong}\\
        Department of Physics 0319, University of California, San Diego\\
        9500 Gilman Drive, La Jolla, CA 92093, USA\\
        E-mail: \email{rickywong@physics.ucsd.edu} }

\abstract{
Extended simulation results and their analysis are reported in
a strongly coupled gauge theory with twelve fermion flavors  
in the fundamental SU(3) color representation. 
The conformality of the model is probed using mass deformed 
conformal finite size scaling (FSS) theory driven by  the fermion mass anomalous dimension. 
Two independent conformal FSS fitting procedures are used in the analysis. The first one deploys physics motivated 
scaling functions, complemented by a second fitting procedure with spline based general B-forms
for the scaling functions.
The results at fixed gauge coupling show unresolved problems with the conformal hypothesis. 
\vskip 0.005 in
}

\FullConference{The 30 International Symposium on Lattice Field Theory - Lattice 2012,\\
		June 24-29, 2012,
		Cairns, Australia}

\begin{document}

\section{Introduction}

The frequently discussed model of twelve fermions in the fundamental representation 
of the SU(3) color gauge group remains controversial with continuing recent efforts 
from several lattice investigations~\cite{Fodor:2011tu,Fodor:2012uu,Kieran:2012,Jin:2012dw,Aoki:2012eq,
Aoki:2012kr,Schaich:2012fr,Hasenfratz:2012fp,Cheng:2011ic,Deuzeman:2012ee,
Deuzeman:2012pv,Deuzeman:2011pa,Miura:2011mc} where more extended references can be 
found for the earlier history of the model. The focus of these lattice studies is to establish conformality or chiral 
symmetry breaking in the bulk phase close to the continuum limit.
At finite cutoff,  two different 
strategies can be used to deal with finite volume dependence while probing the conformal and $\chi{\rm SB}$ hypotheses 
of the bulk phase.
The first strategy extrapolates the spectrum to infinite volume at fixed fermion mass $m$ where the leading 
finite size correction is exponentially small and assumed to be determined by the lowest mass with pion quantum numbers.
From the mass spectrum of the infinite volume extrapolation 
the mass deformed conformal scaling behavior can be probed and compared with $\chi{\rm SB}$ behavior when the fermion mass is varied 
in the infinite volume limit. This limit requires very large volumes  and  it is difficult to reach with existing computational resources.
Probing the conformal hypothesis without intrinsic scale, the second strategy takes full advantage of the conformal
FSS behavior when approaching the $m=0$ critical surface at fixed finite size $L$. 
Different from the first strategy, the finite volume corrections are not exponentially small and a much larger 
data set can be analyzed closer to the critical surface.
Results from the conformal FSS analysis are presented here, significantly extending what was reported earlier~\cite{Fodor:2011tu}.

\section{The algorithm and the data base from the simulations}

We have used the tree-level Symanzik-improved gauge action with $\beta=2.2$  gauge coupling 
and with two levels of stout smearing in the staggered fermion action~\cite{Aoki:2005vt}. 
The standard HMC algorithm was used with multiple time scales and the
Omelyan integrator.
Our error analysis of  hadron masses used correlated fitting with double jackknife procedure on the covariance matrices.
The time histories of the fermion condensate, the plaquette, 
and correlators are used to monitor autocorrelation times in the simulations.

Extending earlier work~\cite{Fodor:2011tu}, we have new simulation results at $\beta=2.2$ in the 
fermion mass range ${\rm m=0.002-0.025}$ at lattice volumes
$20^3\times 40$,  $24^3\times 48$,  $28^3\times 56$, $32^3\times 64$, $40^3\times 80$, and $48^3\times 96$. 
The extended data base now has the  ${\rm m=0.002-0.035}$  fermion mass range with a span close to a factor of twenty.
The  new low fermion mass range  ${\rm m=0.002,0.004,0.006,0.008}$ is used in the conformal FSS analysis
which over the full set  extends the lattice pion correlation length from 2.5 to 20 in the infinite volume limit.
Results from the two lowest masses at ${\rm m=0.002,0.004}$ are not included in the current analysis and will be reported later.
For further control on finite volume dependence,  large $48^3\times 96$ 
runs were continued to two thousand trajectories at ${\rm m=0.01}$ and ${\rm m=0.015}$.  
Four runs were 
further added at $40^3\times 80$ with ${\rm m=0.01, 0.15, 0.02, 0.025}$.
The new and updated data sets were subjected to conformal FSS analysis and $\chi{\rm SB}$ tests of the
$\langle \bar{\psi}\psi\rangle$ chiral condensate. 
Preliminary analysis of the results already appeared a few months ago~\cite{Fodor:2012uu}.

\section{The bulk phase diagram}
The bulk phase structure of the model remains controversial, particularly the critically important weak coupling phase.
In addition to our spectroscopy and conformal FSS runs, we ran extensive scans at various fixed volumes and fixed fermion masses
to explore the bulk phase structure. The bare coupling $\beta$ was varied over a large range starting 
from very small $\beta$ values deep in the strong
coupling regime to the weak coupling phase at $\beta=2.2$ where the conformal and $\chi{\rm SB}$ analyses
were done.
Fermion masses ${\rm m=0.007,0.01, 0.02}$ were used
in the scans with spatial lattice sizes $L=8,12,16,20,24,32$ running a large densely spaced set in the important
and much discussed intermediate region in transit from strong coupling to weak coupling.
These scans were also extended to $N_f=2,4,6,8,10,12,14,16$ flavors. 
\begin{figure}[hbt!]
\begin{center}
\begin{tabular}{cc}
\includegraphics[width=0.39\textwidth]{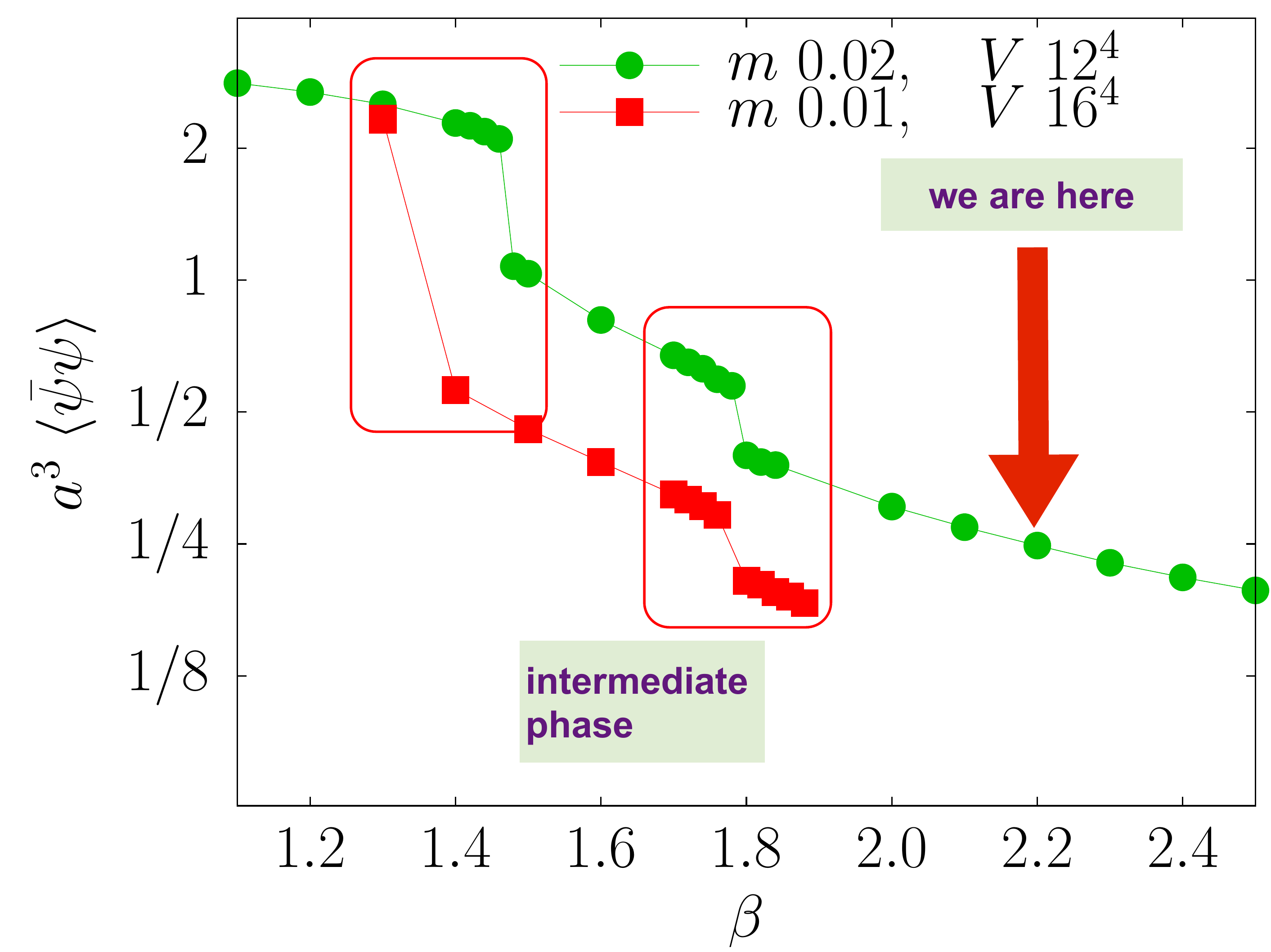}&
 \includegraphics[width=0.35\textwidth]{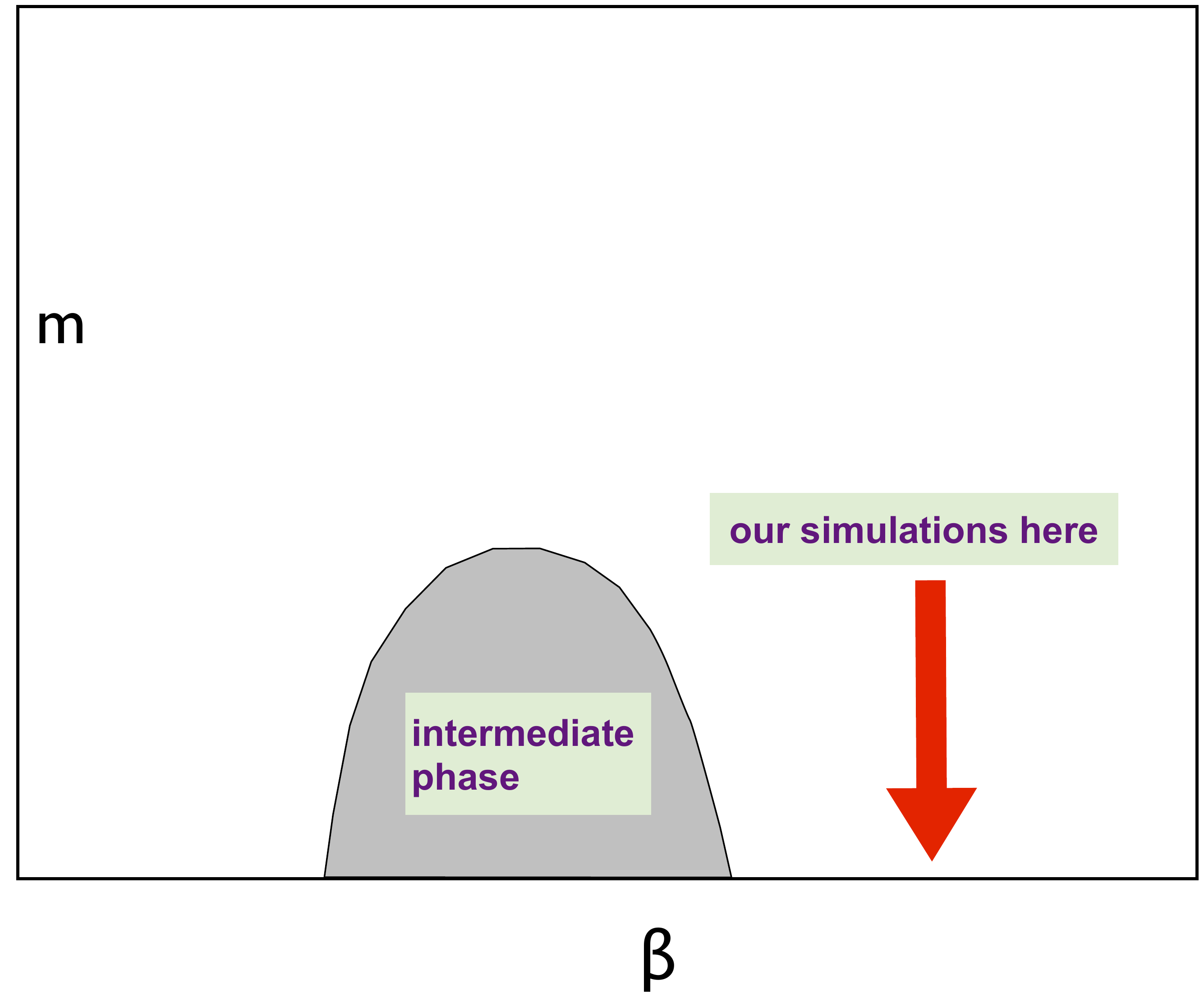}
\end{tabular}
\end{center}
\vskip -0.2in
\caption{\footnotesize On the left, scans of the phase diagram by monitoring the chiral condensate are plotted as a function of  $\beta$ 
at two different fermion masses for $N_f=12$. The schematic bulk phase diagram is sketched on the right, a plot suggested from findings
in~\cite{Cheng:2011ic}.}
\label{fig:PhaseDiagram1}
\end{figure}
Two representative scans of the bulk behavior of the chiral condensate 
$\langle \bar{\psi}\psi\rangle$  are shown  in Figure~\ref{fig:PhaseDiagram1} as we vary $\beta$ from strong to weak coupling.
Three distinct regions emerge at fixed volume and fixed fermion mass showing strong coupling behavior 
for $\beta < 1.4$ with a large chiral condensate, an intermediate phase for
$1.4 < \beta <1.8$ with a sudden drop in $\langle \bar{\psi}\psi\rangle$, and a weak coupling phase for $\beta > 1.8$ with 
a further drop in $\langle \bar{\psi}\psi\rangle$. 
We also observed another distinct feature of the intermediate phase
(first seen in~\cite{Cheng:2011ic}) where the pion correlator exhibits oscillations associated 
with the appearance of a non-degenerate parity partner state. 
This structure does not exist in the weak coupling phase.
A similar pattern of phases was also seen in scans at $N_f=8$. 
Our physics simulations and conformal FSS hypothesis tests were done well inside the weak coupling phase 
at $\beta=2.2$  without influence from the intermediate phase, as indicated in Figure~\ref{fig:PhaseDiagram1}.

A similar phase structure has been observed
independently by three different groups~\cite{Jin:2012dw,Cheng:2011ic,Deuzeman:2011pa}. 
The newfound 
order parameter of broken shift symmetry in the intermediate phase is an interesting additional development 
in the study of the esoteric intermediate phase when HYP smearing is used in the staggered fermion action~\cite{Cheng:2011ic}. 
The intermediate phase only exists in a finite interval of the lattice gauge coupling for
small enough fermion masses, as schematically sketched in Figure~\ref{fig:PhaseDiagram1}. 
The real interest is, of course, in the nature of the weak coupling phase. 
Based on axial U(1) symmetry considerations, arguments were presented in~\cite{Deuzeman:2011pa} 
in favor of conformal symmetry in the 
weak coupling phase. This argument was criticized  in~\cite{Cheng:2011ic} based on new details of the
broken shift symmetry with chiral symmetry restoration they discovered
 at zero temperature in the bulk intermediate phase.

Weak coupling results on the Polyakov loop, the chiral condensate,
and spectroscopy were also presented in ~\cite{Cheng:2011ic} as indications of conformal symmetry in the weak coupling phase.
The blocked Polyakov loop was reported to jump from zero to a large finite value in crossing to the weak 
coupling phase ~\cite{Cheng:2011ic}.
A confining potential was reported  in the intermediate phase with broken shift symmetry which turned into a Coulomb potential 
without a string tension in the weak coupling phase~\cite{Cheng:2011ic}.
It was also asserted that the observed chiral condensate and the related Dirac spectrum show the recovery of exact chiral 
symmetry in the massless fermion limit of the weak coupling phase consistent with observed degeneracies of parity partners
even at finite fermion masses. In conclusion, conformal behavior was suggested for the bulk weak coupling phase.

The results reported in~\cite{Cheng:2011ic} suggesting a chirally symmetric deconfined conformal phase 
are in disagreement with what we found
earlier~\cite{Fodor:2011tu} and in the 
extended new analysis~\cite{Fodor:2012uu,Kieran:2012}. Using lattice volumes several times larger than the simulations in~\cite{Cheng:2011ic} 
we find a vanishing Polyakov loop at zero temperature in the weak coupling phase and a confining 
potential at a pion correlation length which is significantly larger and consequently 
more relevant than the related findings in~\cite{Cheng:2011ic}. We also observe the splitting of parity partners
at finite fermion mass. Consequently, and differing from~\cite{Cheng:2011ic},  our findings in large volumes are consistent 
with a chirally broken weak coupling phase.
We turn now to conformal FSS tests of the conformal hypothesis in the weak coupling phase 
gathering further evidence towards more definitive conclusions.

\section{Conformal finite size scaling analysis}
\begin{figure}[!t]
\begin{center}
\begin{tabular}{cc}
\includegraphics[height=5.5cm]{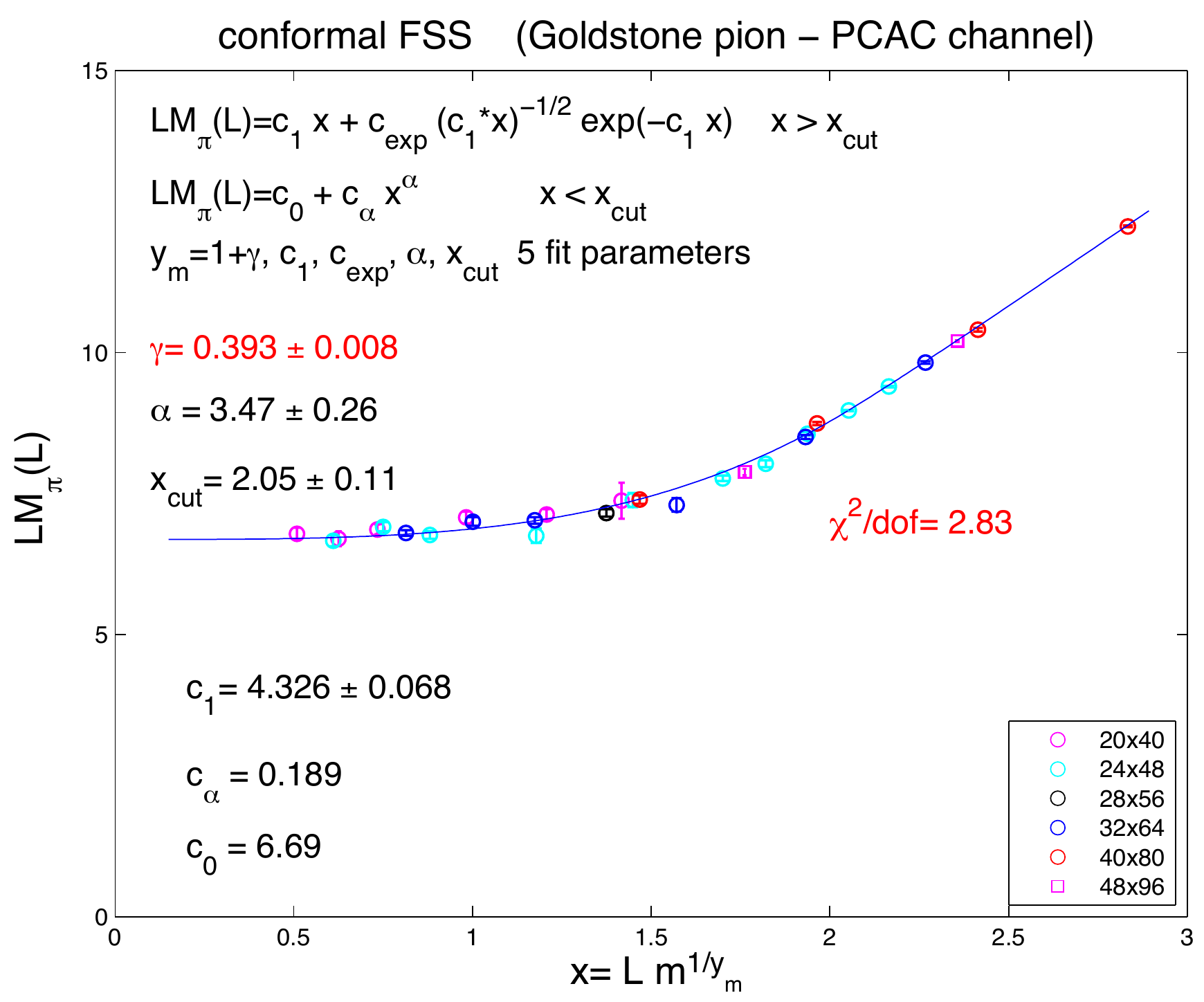}&
\includegraphics[height=5.5cm]{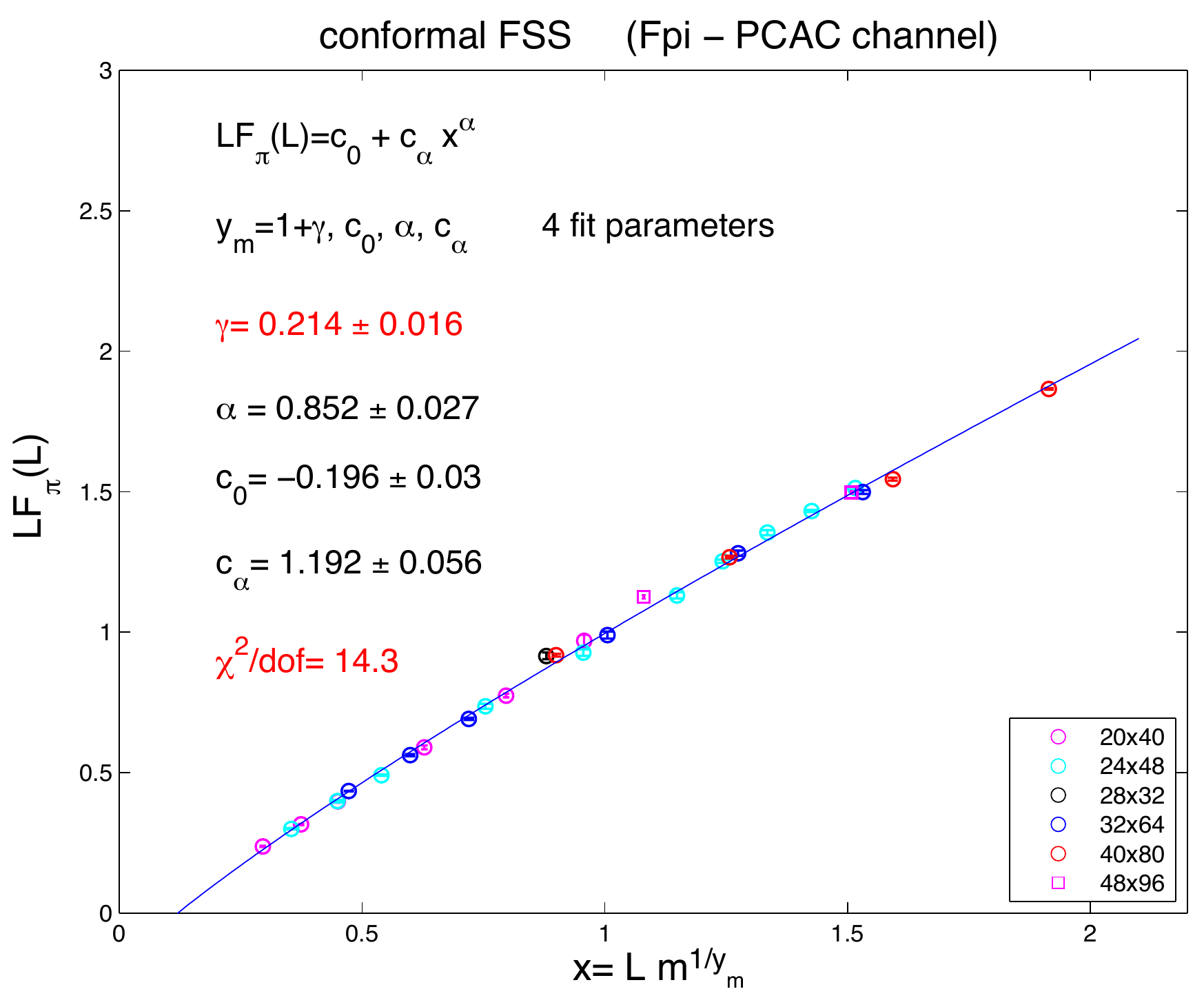}\\
\includegraphics[height=5.5cm]{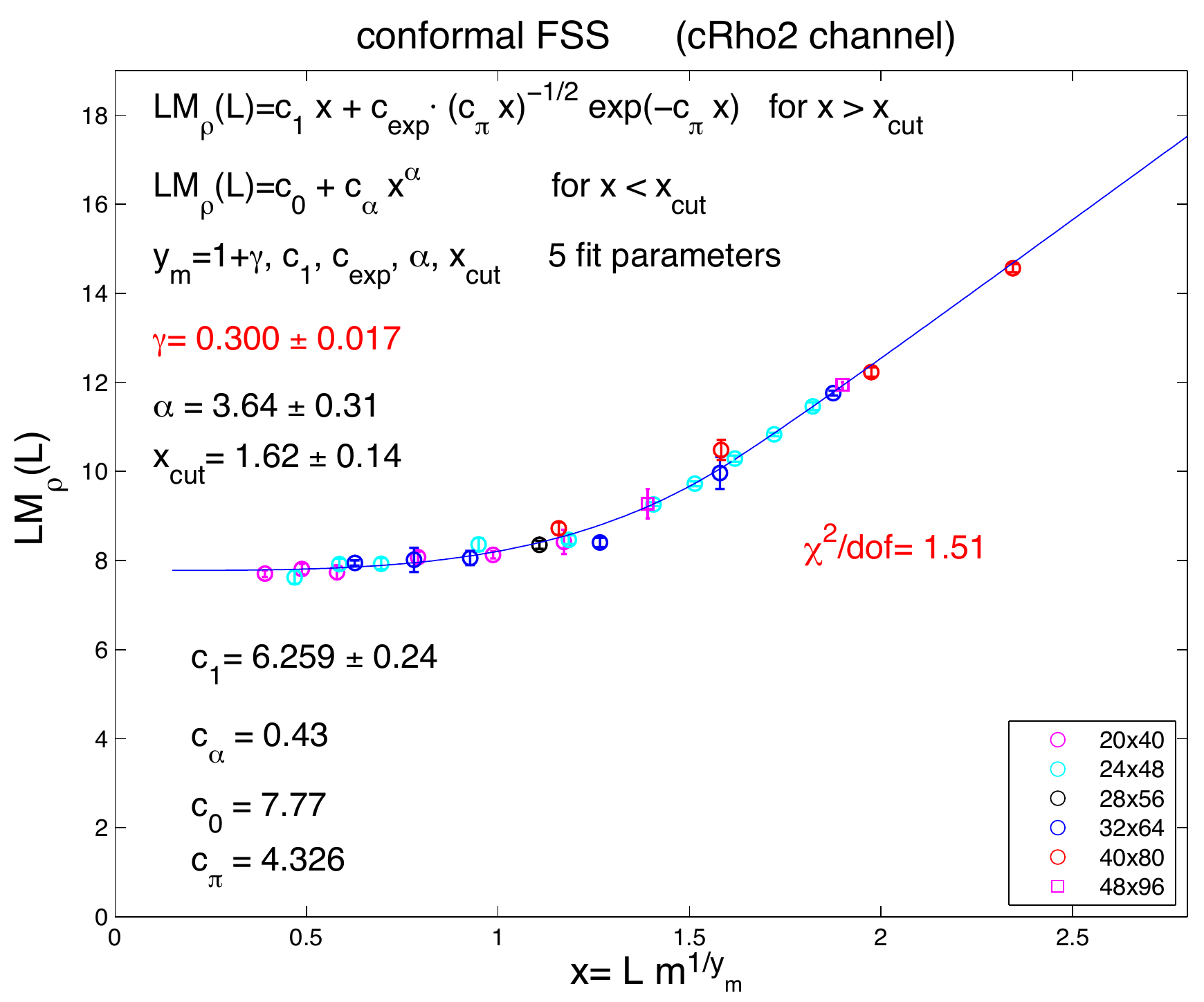}&
\includegraphics[height=5.5cm]{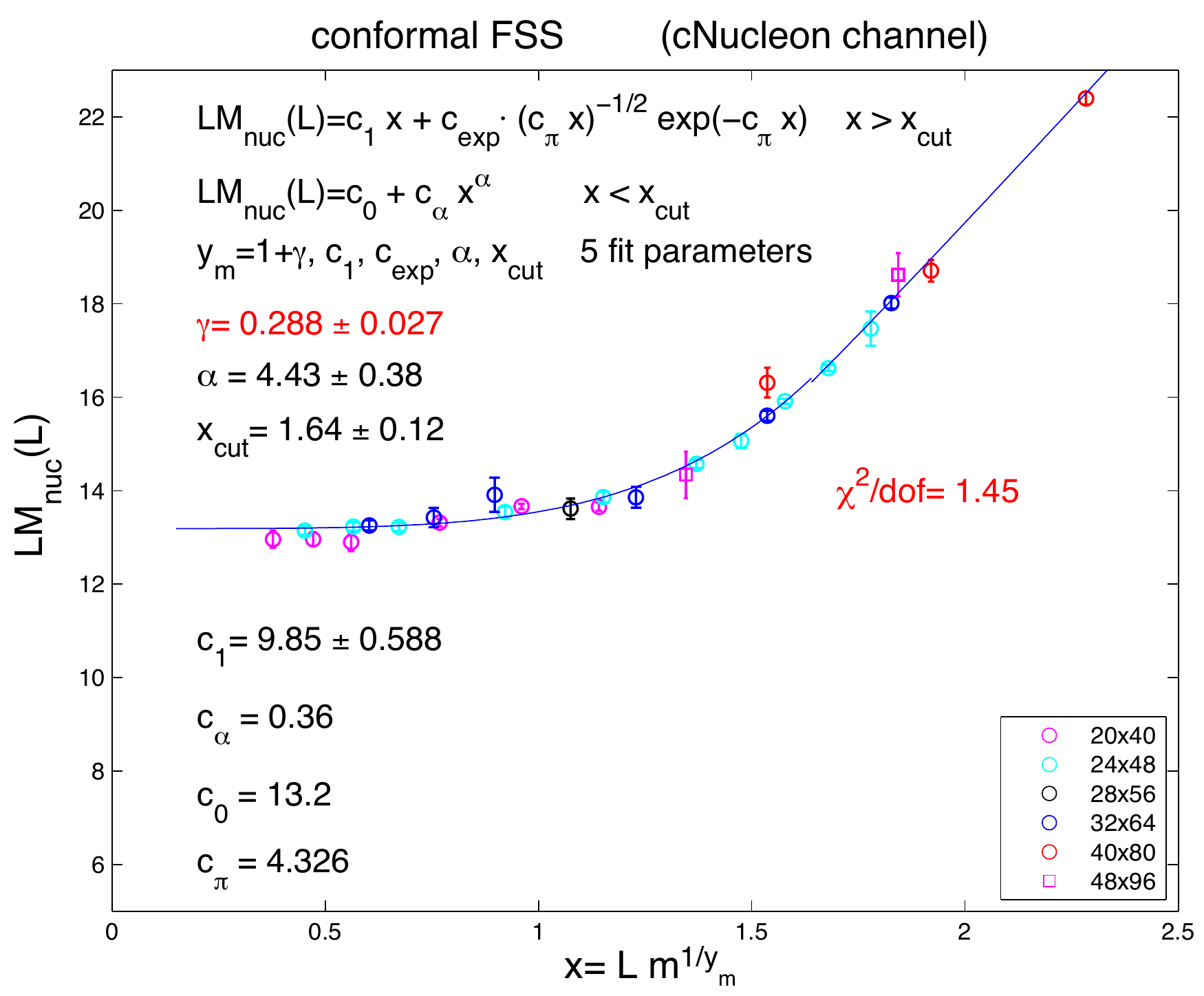}
\end{tabular}
\end{center}
\caption{{  Conformal FSS fits in four different quantum number channels. 
The fits are performed in each channel separately.
Since the $\gamma$ values vary considerably from channel to channel, a simultaneous global fit to the 
combined channels with the same $\gamma$ exponent, as required by conformal FSS theory, is bound to fail.}}
\label{fig:ConformFSS}
\end{figure}
The expected leading FSS  form for any mass $M$ in the spectrum, or for the decay constant $F_\pi$, 
scaled with the linear size $L$ of the spatial volume,
is given by a scaling function $L\cdot M=f(x)$ where $x=L\cdot m^{1/1+\gamma}$ is the conformal 
scaling variable. The scaling form sets in close to the critical surface for small $m$ values.
The scaling functions $f(x)$ can depend on the quantum numbers of the states 
but the scaling variable is expected to have the same form with identical $\gamma$ exponent in 
each quantum number channel~\cite{DelDebbio:2010hx,Bursa:2010xn,DelDebbio:2010ze,DelDebbio:2010jy,DelDebbio:2011kp}. 
In sub-leading order there are conformal FSS scaling violation effects which 
are exhibited as a combined cutoff and $L$-dependent leading  correction with the modified form 
$L\cdot M=f(x) + L^{-\omega}g(x)$ where the scaling correction exponent $\omega$ is determined at the 
infrared fixed point (IRFP) $g^*$
of the $\beta$-function as $\omega = \beta'(g^*)$. This assumes that the mass deformation away from the critical
surface is the only  relevant perturbation around the IRFP. The leading scaling correction term close enough 
to the critical surface dominates any other corrections which are further supressed by additional inverse powers of $L$.
To detect the leading scaling violation effect requires high precision data with fits to
scaling functions $f(x)$ and $g(x)$ and the critical exponent $\omega$. 

\subsection{Conformal FSS fitting procedure with restricted scaling functions}

We applied conformal FSS theory to our data sets in the fermion mass range $m=0.006-0.035$ with lattice sizes ranging 
in the fits from  $20^3\times 40$ to $48^3\times 96$. Two different FSS fitting procedures were applied.
In the first procedure, we defined a scaling function $f(x)$ for each mass M with five independent
fitting parameters. The fitting function $f(x)$ was divided
into two regions separated at  the joint $x=x_{cut}$. Different forms were chosen on the two sides of $x_{cut}$ for
the expected conformal behavior.
For large $x>x_{cut}$, the function $f(x)=c_1 x + c_{exp}(c_1x)^{-1/2}{\rm exp}(-c_1x)$ with
parameters $c_1$ and $c_{exp}$ describes the $L$-independent limit  $M \sim c_1 m^{1/1+\gamma}$ 
at fixed $m$ and $L\rightarrow\infty$.
The $c_{exp}$ amplitude sets the size of the leading small exponential correction from the 
finite volume effect of the lightest Goldstone pion state wrapping around the spatial volume.
Since $f(0)=c_0$ is expected from conformal FSS with some power corrections at small $x$, we applied the simple
ansatz $f(x)=c_0 + c_{\alpha}x^{\alpha}$ for $x<x_{cut}$ (a more general polynomial function in the small $x$ region will not
change the conclusions from the fits).
From the fit to the PCAC Goldstone pion channel the parameter $c_{\pi}=c_1$ was determined and used as input in 
the exponential terms  of the other channels with ${\rm exp(-c_\pi L}$).
The critical exponent $\gamma$ was included among the five fitting parameters, in addition to $c_0,~c_1,~c_{exp}$, and $x_{cut}$.

The composite particle masses in several  quantum number channels can be reasonably fitted 
with  conformal scaling functions $f(x)$ as shown in Figure~\ref{fig:ConformFSS} but the values of the 
critical exponent $\gamma$ are incompatible across different channels. The required global conformal FSS fit will fail
with a single exponent $\gamma$ across all quantum numbers.
In the fits for $F_\pi$ in the PCAC pion channel we only kept four parameters because the asymptotic form with exponentially 
small correction was zero within error. Actually, the data of $F_\pi$ did not allow a successful conformal fit with any shape chosen
for its scaling function $f(x)$ which looks very different from the scaling functions of composite particle masses. The unexpectedly
curious behavior of the $F_{\pi}$ data set against conformal FSS remains unresolved. Our dataset is not accurate enough to successfully
resolve subleading corrections in the fits.

\subsection{Generalized FSS fitting procedure with spline based general B-form}

Following a new  fitting strategy, we investigated if the failed global conformal FSS analysis can 
be attributed to restrictions on the conformal 
scaling functions $f(x)$. The restrictions were  manifest in the physics-motivated fitting procedure we applied above. 
Our new general approach is different from~\cite{Appelquist:2011dp,DeGrand:2011cu} but addresses related issues. 
We developed a general least-squares fitting
procedure to the scaling functions using the B-form of spline functions~\cite{DeBoor:1978} without additional restrictions. 
In this procedure, the function $f(x)$ is described
by piece-wise polynomial forms constructed from spline base functions with general
coefficients in overlapping intervals of the scaling variable $x$ which depends on the exponent $\gamma$. 
The shape of the B-form can be changed without limitations by increasing the number of base functions and the number
of scaling intervals in $x$  bracketing the overlapping data range.
The details of this new analysis will be reported elsewhere.  
\begin{figure}[h]
\begin{center}
\begin{tabular}{ccc}
\includegraphics[height=4cm]{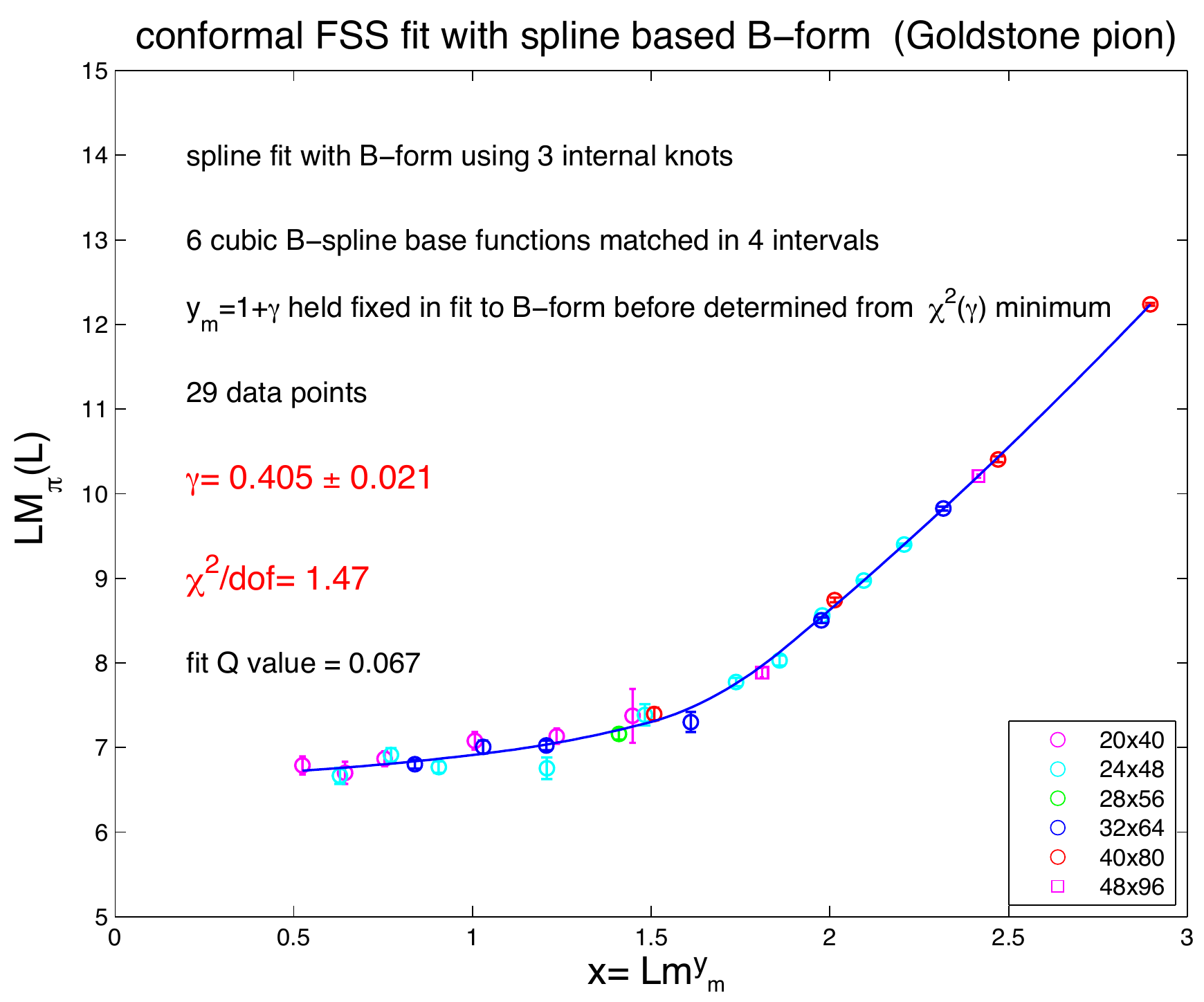}&
\includegraphics[height=4cm]{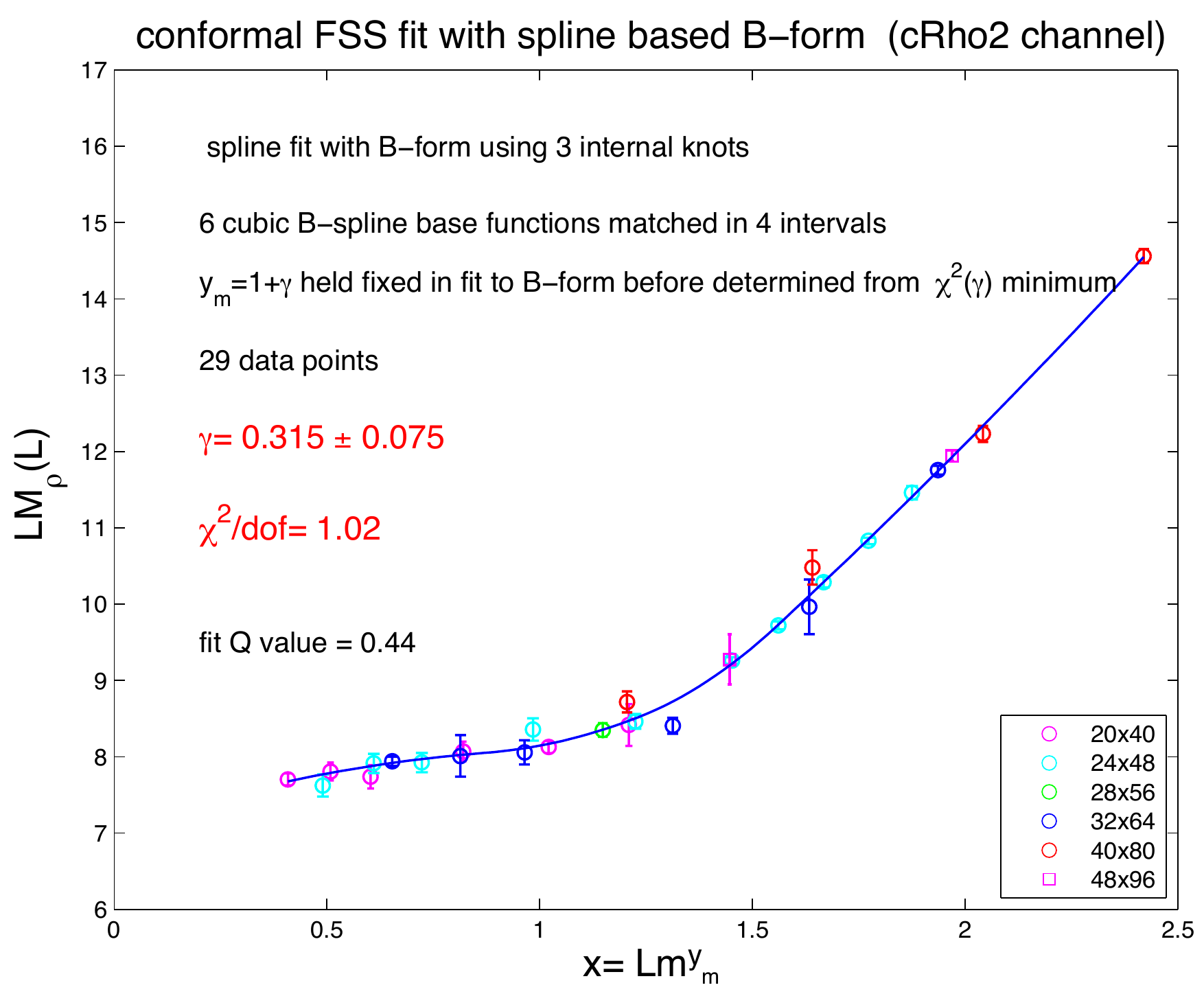}
\includegraphics[height=4cm]{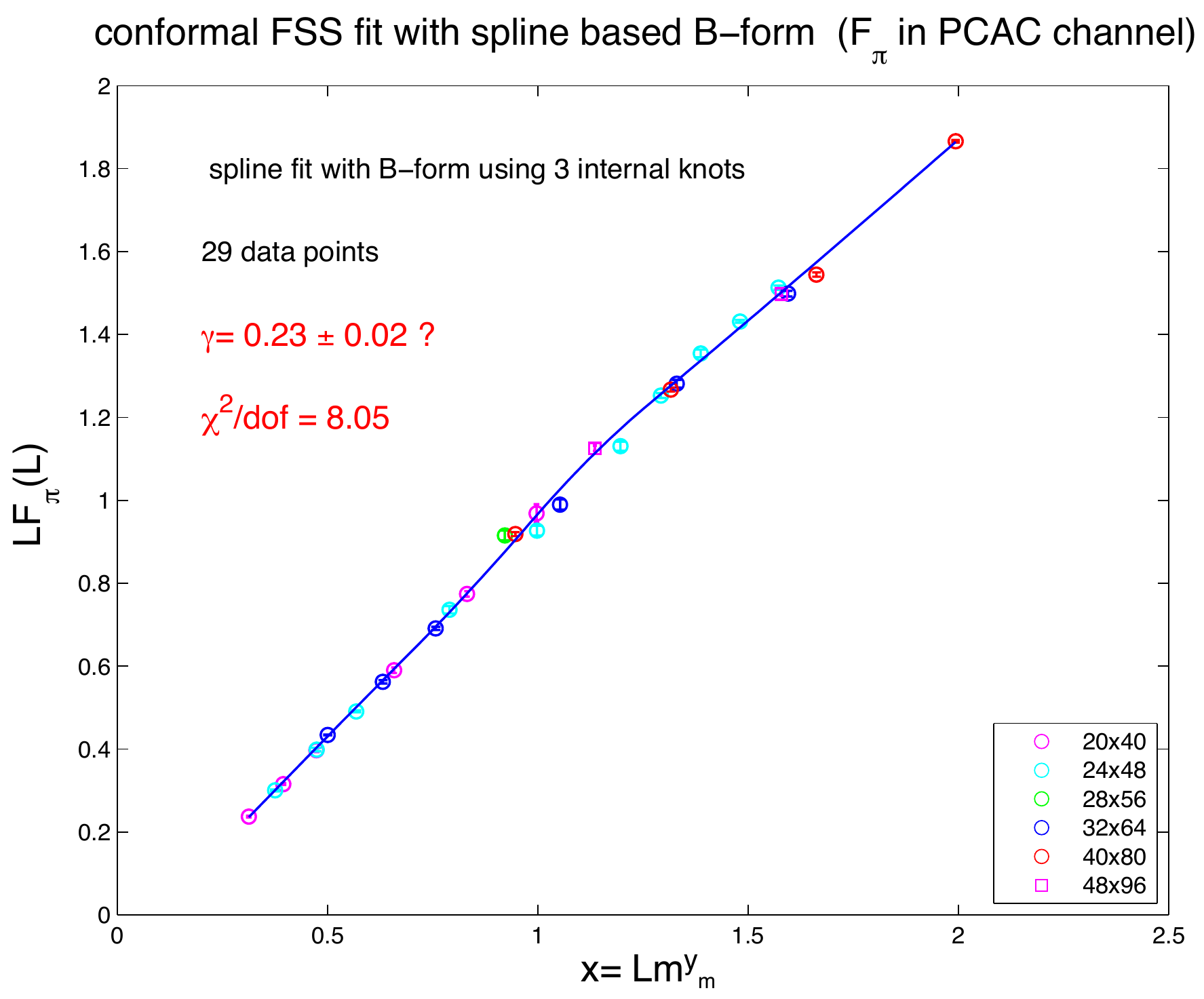}&
\end{tabular}
\end{center}
\caption{{\footnotesize  Conformal FSS fits using spline based B-forms in three different channels. 
The fits are performed in each channel separately with the question mark on $\gamma$ indicating
difficulties of error estimates in bad fits of  $F_\pi$ . }}
\label{fig:splineConformFSS}
\end{figure}

Our fitting procedure in its setup requires two steps.
In the first step, for any fixed choice of the exponent $\gamma$, the 
best fitted function $f(x)$ is determined in spline function B-form from the minimization of the weighted $\chi^2$ expression.
According to a general algorithm, the $x$-range of the data set is divided into intervals separated by internal knots and 
adding end point knots for B-form spline construction.  The number of coefficients is determined by the number of knots and the order of 
the spline polynomials of the sub-intervals. 
The weighted $\chi^2$ sum is minimized with respect to the coefficients of the base functions in the B-form. This will produce
the best fit for fixed $\gamma$ with a minimized $\chi^2$ sum which will depend on $\gamma$.
In the second step, we minimize the $\chi^2$ sum with respect to $\gamma$ to determine the best fit
of the critical exponent. The one-$\sigma$ confidence interval is determined from the variation of the $\chi^2$ sum as a function
of $\gamma$.

In Figure~\ref{fig:splineConformFSS} we show three typical fits for illustration. The fit to the Goldstone pion in the PCAC channel 
improved as expected, with considerable increase in the error. The tension across channels decreased, as illustrated by
comparison with the rho-channel fit, but the fit to $F_{\pi}$ remained unacceptable. 
With the extended data set we are unable to reproduce results in ~\cite{Appelquist:2011dp,DeGrand:2011cu}
which used tables from our earlier limited subset of data~\cite{Fodor:2011tu} to argue in favor of consistency with the conformal phase.
It is important to emphasize that we have not reached 
definitive conclusions about the failure of conformal tests.  As we stated earlier~\cite{Fodor:2011tu}, 
we have not analyzed yet the leading scaling violation effects and did not investigate if the good scaling form in separate 
quantum number channels can be explained in the chirally broken phase by strongly sqeezed wave function effects. 
In disagreement with ~\cite{DeGrand:2011cu}, conformal FSS based analysis of the spectrum and related 
sum rules on moments of the correlators we have been developing are deep renormalization group based probes 
of the conformal phase. 
 As explained in our forthcoming publication, we remain skeptical about the fitting procedure followed in
~\cite{Appelquist:2011dp} with efforts to rescue the conformal interpretation. The issues are not settled and ultimately 
will be decided in more definitive analyses.

\section*{Acknowledgments}
\vskip -0.1in
We acknowledge support by the DOE under grant DE-FG02-90ER40546, by the NSF under grants 0704171 and 0970137, 
by the EU Framework Programme 7 grant (FP7/2007-2013)/ERC No 208740, and 
by the Deutsche Forschungsgemeinschaft grant SFB-TR 55. Computational resources were 
provided by USQCD at Fermilab and JLab, by the NSF XSEDE program,
and by the University of Wuppertal. 
KH wishes to thank the Institute for Theoretical Physics and the Albert Einstein Center for Fundamental Physics at Bern University for their support.
KH and JK wish to thank the Galileo Galilei Institute for Theoretical Physics and INFN for their hospitality and support at the workshop
"New Frontiers in Lattice Gauge Theories".

\end{document}